\documentclass[journal]{article}
\usepackage[pass,a4paper]{geometry}
\usepackage[english]{babel}
\usepackage{color,graphicx}
\usepackage{amsmath,epsfig,amsfonts,dsfont}
\usepackage{cite}
\newcommand{\bX}{\boldsymbol{X}}
\newcommand{\bx}{\boldsymbol{x}}
\newcommand{\ba}{\boldsymbol{a}}

\newcommand{\by}{\boldsymbol{y}}

\newcommand{\bA}{\boldsymbol{A}}
\newcommand{\bB}{\boldsymbol{B}}

\newcommand{\bD}{\boldsymbol{D}}

\newcommand{\bP}{\boldsymbol{P}}
\newcommand{\bW}{\boldsymbol{W}}
\newcommand{\bR}{\boldsymbol{R}}
\newcommand{\R}{\mathds{R}}

\newcommand{\bS}{\boldsymbol{S}}
\newcommand{\bu}{\boldsymbol{u}}
\newcommand{\bv}{\boldsymbol{v}}
\newcommand{\bU}{\boldsymbol{U}}
\newcommand{\bV}{\boldsymbol{V}}
\newcommand{\bh}{\boldsymbol{h}}
\newcommand{\bY}{\boldsymbol{Y}}
\newcommand{\bZ}{\boldsymbol{Z}}
\newcommand{\bL}{\boldsymbol{L}}
\newcommand{\diag}{\textrm{diag}}

\newtheorem{theorem}{Theorem}

\title{Phase oscillator model for noisy oscillators}

\author{Michele Bonnin \thanks{Department of Electronics and Telecommunications, Politecnico di Torino, Turin, Italy (e-mail: michele.bonnin@polito.it).}}

\begin{document}

\maketitle

\abstract{The Kuramoto model has become a paradigm to describe the dynamics of nonlinear oscillator under the influence of external perturbations, both deterministic and stochastic. It is based on the idea to describe the oscillator dynamics by a scalar differential equation, that defines the time evolution for the phase of the oscillator. Starting from a phase and amplitude description of noisy oscillators, we discuss the reduction to a phase oscillator model, analogous to the Kuramoto model. The model derived shows that the phase noise problem is a drift--diffusion process. Even in the case where the expected amplitude remains unchanged, the unavoidable amplitude fluctuations do change the expected frequency, and the frequency shift depends on the amplitude variance. We discuss different degrees of approximation, yielding increasingly accurate phase reduced descriptions of noisy oscillators.
} %end of abstract
\section{Introduction}
\label{intro}

Nonlinear oscillators are key components of many modern electronic devices. They are used in communication systems for modulation and demodulation of data, and in  digital systems to provide a time reference frame to synchronize operations. An ideal oscillator would exhibit a perfectly periodic behavior, represented by a limit cycle in its state space. However, the output of actual oscillators is always corrupted by different types of disturbances, such as internal noise sources, thermal noise, and interactions with the environment. Noise sources in electronic circuits cause diffusion (or jitter) of the oscillator phase, ultimately inducing phase noise and time jitter \cite{Lax1967,Kaertner1990}. In the frequency domain, the consequences are a broadening and a shift in the power spectrum peaks, that may eventually produce
interference on nearby frequency bands in transmitters and receivers \cite{Demir2000,Bonnin2013}.

Phase noise is a typical time domain phenomenon arising from the random walk process that the phase undergoes \cite{Maffezzoni2012}. Therefore, time domain models are particularly well suited for the analysis of phase noise. The Kuramoto model has become a paradigm to describe the dynamics of nonlinear oscillator under the influence of external perturbations, both deterministic and stochastic \cite{Kuramoto1984,Acebron2005}. Kuramoto model is based on the idea to describe the state of a nonlinear oscillator by a single scalar variable, the phase, that represents the position of the system on a reference orbit. To derive the model, the solution of the perturbed system is written as a time shifted version of the unperturbed one, and an equation is derived for the time shift. The time shift is interpreted as a phase shift normalized by the free running frequency, and the equation for the phase shift is called \emph{phase oscillator}. The amplitude dynamics is completely neglected, under the hypothesis that the limit cycle is strongly stable and that amplitude deviations are immediately absorbed. This assumption is reasonable for most electronic oscillators, but in other cases, e.g. biological oscillators, it is more taken for mathematical convenience than being physically plausible.

Phase oscillator models have become a paradigm to describe the synchronization of an oscillator with an external periodic signal \cite{Lai2004,Maffezzoni2008,Nagashima2014}, phase locking of coupled oscillators \cite{Hoppensteadt2001,Bonnin2008,Maffezzoni2010}, and to investigate the phase noise in oscillators \cite{Kaertner1990,Demir2000,Djurhuus2009}. However, circuits and systems are usually described in terms of state variables and state equations, and the problem of deriving the phase oscillator model starting from the state equations arises. The exact phase oscillator model can be found only for few, trivial nonlinear oscillators. In all other cases, e.g. phase response curves, one must resort to numerical methods that are either approximate in nature, or unsuitable for oscillators of order higher than the second \cite{Izhikevich2007,Guillamon2009}. In the seminal works \cite{Kaertner1990,Demir2000}, the authors derived a phase oscillator model for nonlinear oscillators subject to weak white Gaussian noise. It was shown that, to the first order, the phase noise problem is a diffusion process. In \cite{Yoshimura2008,Bonnin2014} it was shown that phase noise is best described as a drift--diffusion process, i.e. noise also produces a shift in the oscillation frequency. In \cite{Bonnin2016,Bonnin2017} a rigorous description of noisy oscillators is derived, in terms of phase and amplitude deviation. The resulting equations are exact, they are not restricted to the weak noise limit and they are valid for oscillators of any order. The phase and amplitude equations represent the ideal starting point for the derivation of phase oscillator models.

This paper describes a phase oscillator model for nonlinear oscillators subject to white Gaussian noise, analogous to the celebrated Kuramoto model. It is shown that in a neighborhood of the limit cycle, the phase variable introduced in the proposed model coincide with the phase function defined using the concept of isochrons. The novel phase oscillator model gives better approximations of the full dynamics than other reduced models previously proposed in literature. The relationships with these previous models are discussed. In particular, it is shown that previous models can be derived from the novel one, introducing increasing order of approximation. As an example, the theory is applied to Duffing--van der Pol system with modulated noise.

\section{Phase and amplitude deviation equations for noisy oscillators}\label{noisy oscillators}
Different types of random disturbances in electronics and biology can be modeled as white Gaussian noise. As a consequence electronics and biological oscillators with noise can be conveniently described by stochastic differential equation (SDEs) \cite{Oksendal2003}
\begin{equation}
d \bX_t = \ba(\bX_t) \, d t+ \varepsilon \bB(\bX_t) \, d \bW_t \label{sec2-eq1}
\end{equation}
where $\bX_t: \R \mapsto \R^n$ is a stochastic process describing the state of the oscillator, $\ba: \R^n \mapsto \R^n$ is a vector valued function that defines the oscillator dynamics, $\bB : \R^n \mapsto \R^{n,m}$ is a real valued $n \times m$ matrix, $\varepsilon$ is a parameter that measures the noise intensity and $\bW_t : \R \mapsto \R^m$ is a vector of Brownian motion components (the integral of a white noise). We shall assume $\ba \in \mathcal{C}^k(\Omega \subseteq \R^n)$, with $k\ge 1$ and that $\bB$ satisfy a Lipschitz condition, to guarantee the existence and uniqueness of the solution \cite{Oksendal2003}. Following \cite{Kaertner1990}, we shall interpret \eqref{sec2-eq1} as an It\^o SDE.

In absence of noise (for $\varepsilon=0$) eq. \eqref{sec2-eq1} reduces to a system of ordinary differential equations (ODEs). We assume that these ODEs admit an asymptotically stable $T$--periodic limit cycle denoted by $\bx_s(t)$. We define the unit vector tangent to the limit cycle
\begin{equation}
\bu_1(t) = \dfrac{\ba(\bx_s(t))}{|\ba(\bx_s(t))|} \label{sec2-eq2}
\end{equation}
Together with $\bu_1(t)$ we consider other $n-1$ linear independent vectors $\bu_2(t),\ldots,\bu_n(t)$, such that the set $\{\bu_1(t),\ldots,\bu_n(t)\}$ is a basis for $\R^n$, for all $t$. Given the matrix $\bU(t) = [\bu_1(t),\ldots,\bu_n(t)]$, we define the reciprocal vectors $\bv_1^T(t),\ldots,\bv_n^T(t)$ to be the rows of the inverse matrix $\bV(t) = \bU^{-1}(t)$. Thus $\{\bv_1(t),\ldots,\bv_n(t)\}$ also span $\R^n$ and the bi--orthogonality condition $\bv_i^T \bu_j = \bu_i^T \bv_j = \delta_{ij}$ for all $t$, holds. We shall also use the matrices $\bY(t) = [\bu_2(t),\ldots,\bu_n(t)]$, $\bZ(t) = [\bv_2(t),\ldots,\bv_n(t)]$, and the modulus of the vector field evaluated on the limit cycle, $r(t) = |\ba(\bx_s(t))|$.

The phase of a nonlinear oscillator can be interpreted as an elapsed time from an initial reference point. Thus, the phase represents a new parametrization of the trajectories. Together with the phase function $\theta: \R \mapsto \R$ we shall consider an amplitude deviation function $\bR : \R \mapsto \R^{n-1}$, with  $\theta, \bR \in \mathcal{C}^m(\R)$, $m\ge2$. The amplitude deviation $\bR(\bx(t))$ is interpreted as an orbital deviation from the limit cycle. In \cite{Bonnin2016,Bonnin2017} the following theorem was established, that gives the phase and amplitude deviation equations corresponding to the noisy oscillator \eqref{sec2-eq1}

\begin{theorem}\label{theorem1}
Consider the It\^o SDEs \eqref{sec2-eq1} such that the ODEs obtained setting $\varepsilon=0$ admit a $T$--periodic limit cycle $\bx_s(t)$. Let $\{\bu_1(t),\ldots,\bu_n(t)\}$ and $\{\bv_1(t),\ldots,\bv_n(t)\}$ be two reciprocal bases such that $\bu_1(t)$ is given by \eqref{sec2-eq2} and such that the bi--orthogonality condition $\bv_i^T \bu_j = \bu_i^T \bv_j = \delta_{ij}$ holds. Consider the coordinate transformation
\begin{equation}
\bx = \bh(\theta,\bR) = \bx_s(\theta(t)) + \bY(\theta(t)) \, \bR(t) \label{sec2-eq3}
\end{equation}
Then a neighborhood of the limit cycle $\bx_s(t)$ exists, where the phase $\theta(t)$ and the amplitude $\bR(t)$ are It\^o processes and satisfy
\begin{subequations}
\begin{align}
d\theta = & \big[1 + a_{\theta}(\theta,\bR) + \varepsilon^2 \, \hat a_{\theta}(\theta,\bR) \big] dt + \varepsilon \bB_{\theta}(\theta,\bR) \, d\bW_t \label{sec2-eq4a} \\[2ex]
d \bR = & \big[\bL(\theta) \bR + \ba_{\bR}(\theta,\bR) + \varepsilon^2 \hat \ba_{\bR}(\theta,\bR) \big] dt  + \varepsilon \bB_{\bR}(\theta,\bR) \, d\bW_t \label{sec2-eq4b}
\end{align}\label{sec2-eq4}
\end{subequations}
where (explicit dependence on $\theta$ and $t$ is omitted for simplicity)
\begin{subequations}
\begin{align}
a_{\theta}(\theta,\bR)  = & \left(r + \bv_1^T\dfrac{\partial \bY}{\partial \theta} \bR \right)^{-1} \bv_1^T \bigg[\ba(\bx_s+\bY \bR) - \ba(\bx_s) -\dfrac{\partial \bY}{\partial \theta} \bR \bigg] \label{sec2-eq5a} \\[1ex]
%%%%%%%%%%%%%%%%%%%%%%%%%%%%%%%%%%%%%%%%%%%%%%%%%%%%%%%%%%%%%%%%%%%%%%%%%%%%%%
\hat a_{\theta}(\theta,\bR) = &  - \left(r + \bv_1^T\dfrac{\partial \bY}{\partial \theta} \bR \right)^{-1} \bv_1^T  \bigg[ \dfrac{\partial \bY}{\partial \theta} \bB_{\bR} \bB_{\theta}^T + \dfrac{1}{2} \bigg( \dfrac{\partial \ba(\bx_s)}{\partial \theta} + \dfrac{\partial^2 \bY}{\partial \theta^2} \, \bR \bigg) \bB_{\theta} \bB_{\theta}^T \bigg]\label{sec2-eq5b} \\[1ex]
%%%%%%%%%%%%%%%%%%%%%%%%%%%%%%%%%%%%%%%%%%%%%%%%%%%%%%%%%%%%%%%%%%%%%%%%%%%%%%%
\bB_{\theta}(\theta,\bR) = & \bigg(r + \bv_1^T\dfrac{\partial \bY}{\partial \theta} \bR \bigg)^{-1}  \bv_1^T \, \bB(\bx_s + \bY \bR) \label{sec2-eq5c} \\[1ex]
%%%%%%%%%%%%%%%%%%%%%%%%%%%%%%%%%%%%%%%%%%%%%%%%%%%%%%%%%%%%%%%%%%%%%%%%%%%%%%%%
\bL(\theta) = & - \bZ^T \dfrac{\partial \bY}{\partial \theta} \label{sec2-eq5d} \\[1ex]
%%%%%%%%%%%%%%%%%%%%%%%%%%%%%%%%%%%%%%%%%%%%%%%%%%%%%%%%%%%%%%%%%%%%%%%%%%%%%%%%
\ba_{\bR}(\theta,\bR) = & - \bZ^T \bigg[ \dfrac{\partial \bY}{\partial \theta} \bR \, a_{\theta} - \ba(\bx_s + \bY \bR) \bigg] \label{sec2-eq5e}\\[1ex]
%%%%%%%%%%%%%%%%%%%%%%%%%%%%%%%%%%%%%%%%%%%%%%%%%%%%%%%%%%%%%%%%%%%%%%%%%%%%%%%%
\hat \ba_{\bR}(\theta,\bR) = &  -\bZ^T \bigg[ \dfrac{\partial \bY}{\partial \theta} \bR \, \hat a_{\theta} + \dfrac{\partial \bY}{\partial \theta} \bB_{\bR} \bB_{\theta}^T + \dfrac{1}{2} \bigg( \dfrac{\partial \ba(\bx_s)}{\partial \theta} + \dfrac{\partial^2 \bY}{\partial \theta^2} \, \bR \bigg) \bB_{\theta} \bB_{\theta}^T  \bigg] \label{sec2-eq5f}\\[1ex]
%%%%%%%%%%%%%%%%%%%%%%%%%%%%%%%%%%%%%%%%%%%%%%%%%%%%%%%%%%%%%%%%%%%%%%%%%%%%%%%%
\bB_{\bR}(\theta,\bR) = & \bZ^T\bB(\bx_s+\bY \bR) - \bZ^T \dfrac{\partial \bY}{\partial \theta} \bR \, \bB_{\theta}(\bx_s + \bY \bR) \label{sec2-eq5g}
\end{align}\label{sec2-eq5}
\end{subequations}
\end{theorem}
\emph{Proof}: See \cite{Bonnin2016,Bonnin2017}.\\

The phase and amplitude equations given in theorem \ref{theorem1} are exact, because no approximation is used in their derivation, and their validity is not limited to weak noise. However, the resulting phase and amplitude depend on the choice of the basis $\{\bu_1,\ldots,\bu_n\}$. The phase of a nonlinear oscillator can be unambiguously defined using the concept of isochrons \cite{Djurhuus2009,Guckenheimer1975,Bonnin2014}. Isochrons are $(n-1)$--dimensional manifolds transverse to the limit cycle. Given a reference point $\bx_s(0)$ on the limit cycle, the isochron transverse to the cycle at $\bx_s(0)$ is
\begin{equation}
 I_{\bx_s(0)} = \left\{ \bx(0) \in \R^n/\bx_s \; : \;
\lim_{t\rightarrow +\infty} ||\bx(t) - \bx_s(t) || = 0 \right\} \label{sec2-eq6}
\end{equation}
That is, $I_{\bx_s(0)}$ is the set of initial conditions $\bx(0)$ such that the trajectories leaving from $\bx(0)$ meet asymptotically on the limit cycle at $\bx_s(t)$. The phase of a nonlinear oscillator is introduced by assigning the same phase to all points on the same isochron, i.e. isochrons are the level sets of the phase function. The following theorem, given in \cite{Bonnin2016,Bonnin2017}, shows that if the Floquet's basis is used to derive eqs. \eqref{sec2-eq4}, the resulting phase locally coincides in a neighborhood of the limit cycle, with the phase defined through isochrons. It was also shown that the resulting phase dynamics can be partially decoupled from the amplitude deviation dynamics.
\begin{theorem}\label{theorem2}
Consider the variational equation for the noiseless ($\varepsilon=0$) oscillator \eqref{sec2-eq1}
\begin{equation}
\dfrac{d\by(t)}{dt} = \bA(t) \, \by(t) \label{sec2-eq7}
\end{equation}
where $\bA(t) = \frac{\partial \ba(\bx_s(t))}{\partial \bx}$ is the Jacobian matrix evaluated on $\bx_s(t)$. Let $\boldsymbol{\Phi}(t) = \bP(t) e^{\bD \,t} \bS_0$ be the fundamental matrix solution of \eqref{sec2-eq7}, where $\bP(t)$ is a $T$--periodic matrix, $\bS_0 = \bP^{-1}(0)$, and $\bD = \diag[\nu_1,\ldots,\nu_n]$ is a diagonal matrix whose diagonal entries are the Floquet's characteristic exponents. If the basis vectors $\{\bu_1(t),\ldots,\bu_n(t)\}$ are chosen such that
\[ \big[ r(t) \bu_1(t),\bu_2(t),\ldots, \bu_n(t)\big] = \bP(t)\]
then, the It\^o processes for the phase and amplitude reduce to
\begin{subequations}
\begin{align}
d\theta  = & \big(1 + \widetilde a_{\theta}(\theta,\bR) + \varepsilon^2 \hat a_{\theta}(\theta,\bR)\big) dt + \varepsilon \bB_{\theta}(\theta,\bR) \, d\bW_t \label{sec3-eq11a} \\[2ex]
d \bR = & \big( \widetilde \bD \, \bR + \widetilde \ba_{\bR}(\theta,\bR) + \varepsilon^2 \hat \ba_{\bR}(\theta,\bR) \big) d\, t + \varepsilon \bB_{\bR}(\theta,\bR) \, d\bW_t \label{sec3-eq11b}
\end{align}
\end{subequations}
where $\widetilde \bD = \diag[\nu_2,\ldots,\nu_n]$, and the Taylor series of $\widetilde a_{\theta}(\theta,\bR)$, $\widetilde \ba_{\bR}(\theta,\bR)$, do not contain linear terms in $\bR$.
\end{theorem}
\emph{Proof}: See \cite{Bonnin2016}.\\

Theorem \ref{theorem2} shows that, if Floquet basis is used, then up to the first order, the phase equation \eqref{sec3-eq11a} is independent on the amplitude. In this sense Floquet basis represents a privileged basis. The relationship between the phase $\theta$ and the phase defined through isochrons can be inferred from \eqref{sec2-eq3}. It is well known that at any time instant the Floquet vectors $\bu_2(t),\ldots,\bu_n(t)$ span the hyperplane tangent to the isochron \cite{Kuramoto1984,Djurhuus2009,Bonnin2014}. For small deviations, in the neighborhood of the limit cycle, isochrons can be approximated by the hyperplane spanned by  $\bu_2(t),\ldots,\bu_n(t)$. Therefore, any point lying on the isochron can be approximated by \eqref{sec2-eq3}.

\section{Phase oscillator model of noisy oscillators}\label{phase reduced model}

A simple phase oscillator model can be derived from \eqref{sec2-eq4}, by substituting to the amplitude deviation the noiseless value $\bR=\mathbf 0$. Such an assumption is made under the hypothesis that amplitude fluctuations instantaneously relax to the stable limit cycle. Taking into account that $a_{\theta}(\theta,\mathbf 0) =0$, and if $\mathcal{O}(\varepsilon^2)$ terms are neglected, the following phase oscillator model, equivalent to those presented in \cite{Kaertner1990,Demir2000}\footnote{Please note that in the present work the angular frequency is normalized by the oscillator's free running frequency}, is obtained
\begin{equation}
d\theta  = dt + \varepsilon \bB_1(\theta,\mathbf{0}) \, d\bW_t \label{sec3-eq1}
\end{equation}
The expect angular frequency is readily found from eq. \eqref{sec3-eq1}. Taking the stochastic expectation on both sides of \eqref{sec3-eq1} and using the zero expectation property of It\^o integral, it follows that $E[d\theta/dt] =1$. However, the prediction that the noise does not modify the expected angular frequency of the oscillator turns out to be incorrect. Retaining $\mathcal{O}(\varepsilon^2)$ terms yields the more accurate model
\begin{equation}
d\theta  =  \big(1 + \varepsilon^2 \hat a_1(\theta,\mathbf{0})\big) dt + \varepsilon \bB_1(\theta,\mathbf{0}) \, d\bW_t \label{sec3-eq2}
\end{equation}
that correspond to those given in \cite{Yoshimura2008,Bonnin2013,Bonnin2014}. The resulting expected angular frequency is $E[d\theta/dt] = 1 + \varepsilon^2 E[\hat a_1(\theta,\mathbf{0})]$. Eq. \eqref{sec3-eq2} gives better predictions than eq. \eqref{sec3-eq1} for moderate values of $\varepsilon$, but the effect of amplitude deviations is still completely neglected. We shall now derive an improved phase oscillator model that takes into account the effect of noise induced amplitude fluctuations on the phase, yet remaining reasonably simple.

%Starting from a SDE, a Fokker--Planck equation (FPE) for the probability density function can be derived \cite{gardiner1985}.
Let $p(\theta,\bR,t)$ be the probability density function (PDF) for the solution of the phase--amplitude model \eqref{sec2-eq4}. The time evolution of the PDF is described by the Fokker--Planck Equation (FPE) given by \cite{Gardiner1985}
\begin{equation} \begin{array}{rl}
\dfrac{\partial p}{\partial t}  = & - \dfrac{\partial }{\partial \theta} \bigg\{  \Big[1 + a_{\theta}(\theta,\bR) + \varepsilon^2 \hat a_{\theta}(\theta,\bR) \Big] p \bigg\}\\
&  \hspace{-10mm}- \displaystyle{\sum_{i=1}^{n-1}} \dfrac{\partial }{\partial R_i} \bigg\{  \Big[\sum_{j=1}^{n-1} L_{ij}(\theta) R_j + a_{R_i}(\theta,\bR)+ \varepsilon^2 \hat a_{R_i}(\theta,\bR) \Big] p \bigg\} + \dfrac{\varepsilon^2}{2} \dfrac{\partial^2}{\partial \theta^2} \Big[ \bB_{\theta}(\theta,\bR) \bB_{\theta}^T(\theta,\bR) \, p \Big]\\ [1ex]
& + \varepsilon^2  \displaystyle \sum_{i=1}^{n-1} \dfrac{\partial^2}{\partial \theta \partial R_i} \Big[ \bB_{\theta}(\theta,\bR) \bB_{R_i}^T(\theta,\bR) \, p \Big]  + \dfrac{\varepsilon^2}{2} \displaystyle{\sum_{i,j=1}^{n-1}} \dfrac{\partial^2}{\partial R_i \partial R_j} \Big[ \bB_{R_i}(\theta,\bR) \bB_{R_j}^T(\theta,\bR) \, p \Big] \label{sec3-eq3}
\end{array} \end{equation}
where $R_i$ denotes the $i^{\textrm{th}}$ component of $\bR$, $L_{ij}$ is the $i,j$ element of the matrix $\bL$, $a_{R_i}$ is the $i^{\textrm{th}}$ component of $\ba_{\bR}$, and $\bB_{R_i}$ is the $i^{\textrm{th}}$ row of matrix $\bB_{\bR}$.\\
We write the amplitude deviation as a small deviation from the noiseless value in the form $\bR= \varepsilon \widetilde \bR$, and we expand the coefficients of \eqref{sec3-eq3} in Taylor series. Taking into account that $a_{\theta}(\theta,\mathbf 0) =0$, $\ba_{\bR}(\theta,\mathbf0) = \mathbf 0$ and that $\frac{\partial}{\partial R_i} = \frac{\partial}{\varepsilon \partial \widetilde R_i}$, the FPE \eqref{sec3-eq3} can be rewritten in compact form
\begin{equation}
\dfrac{\partial p}{\partial t} = (L_0 + \varepsilon L_1 + \varepsilon^2 L_2) p \label{sec3-eq4}
\end{equation}
with operators
\begin{subequations}
\begin{align}
L_0 p & = -\dfrac{\partial p}{\partial \theta} - \sum_{i=1}^{n-1} \dfrac{\partial}{\partial \widetilde R_i}\left[ \sum_{j=1}^{n-1} \left( L_{ij}(\theta) + \dfrac{\partial a_{R_i}}{\partial R_j} \right) \widetilde R_j \,p \right] + \dfrac{1}{2} \sum_{i,j=1}^{n-1} \dfrac{\partial^2}{\partial \widetilde R_i \partial \widetilde R_j} \Big( \bB_{R_i} \bB_{R_j}^T p\Big) \label{sec3-eq5a}\\[1ex]
\nonumber L_1 p & = -\dfrac{\partial}{\partial \theta} \sum_{i=1}^n \left( \dfrac{\partial a_{\theta}}{\partial R_i} \widetilde R_i \,p  \right) - \sum_{i=1}^{n-1} \dfrac{\partial}{\partial \widetilde R_i} \left[ \left( \dfrac{1}{2} \sum_{j,k=1}^{n-1} \dfrac{\partial^2 a_{R_i}}{\partial R_j \partial R_k} \, \widetilde R_j \, \widetilde R_k + \hat a_{R_i} \right) p \right]\\
& + \sum_{i=1}^{n-1} \dfrac{\partial^2}{\partial \theta \partial \widetilde R_i} \left( \bB_{\theta} \bB_{R_i}^T p \right) \label{sec3-eq5b}\\[1ex]
L_2 p & = -\dfrac{\partial}{\partial \theta} \left[ \left( \dfrac{1}{2} \sum_{i,j=1}^{n-1} \dfrac{\partial^2 a_{\theta}}{\partial R_i \partial R_j} \, \widetilde R_i \, \widetilde R_j + \hat a_{\theta} \right) p \right] + \dfrac{1}{2} \dfrac{\partial^2}{\partial \theta^2} \left( \bB_{\theta} \bB_{\theta}^T p \right) \label{sec3-eq5c}
\end{align}\label{sec3-eq5}
\end{subequations}
In eqs. \eqref{sec3-eq5}, the functions $a_{\theta}$, $\ba_{R}$, $\hat a_{\theta}$, $\hat \ba_{\bR}$, $\bB_{\theta}$, $\bB_{\bR}$ and their derivatives are evaluated at $(\theta,\mathbf 0)$.
The zeroth order equation $\frac{\partial p}{\partial t} = L_0 p$ is the FPE corresponding to the SDEs
\begin{subequations}
\begin{align}
d\theta & = dt \label{sec3-eq6a}\\
d \widetilde \bR & = \left[ \bL(\theta) + \dfrac{\partial \ba_{\bR}(\theta,\mathbf 0)}{\partial \bR} \right] \widetilde \bR \, dt + \bB_{\bR}(\theta,\mathbf 0) d \bW_t \label{sec3-eq6b}
\end{align}
\end{subequations}
Eq. \eqref{sec3-eq6a} implies $\theta=t$, while \eqref{sec3-eq6b} is a time dependent Ornstein--Uhlenbeck process. It stems that $\widetilde \bR(\theta)$ is a vector with Gaussian distributed components, and the moments are obtained solving
\begin{subequations}
\begin{align}
\dfrac{d \boldsymbol \mu}{d \theta} & = \left[ \bL(\theta) + \dfrac{\partial \ba_{\bR}}{\partial \bR} \right] \boldsymbol \mu \label{sec3-eq7a}\\[1ex]
\dfrac{d \boldsymbol \sigma}{d \theta} & = \left[ \bL(\theta) + \dfrac{\partial \ba_{\bR}}{\partial \bR} \right] \boldsymbol \sigma + \boldsymbol \sigma \left[ \bL(\theta) + \dfrac{\partial \ba_{\bR}}{\partial \bR} \right]^T + \bB_{\bR} \bB_{\bR}^T \label{sec3-eq7b}
\end{align}\label{sec3-eq7}
\end{subequations}
Without loss of generality we can assume null initial conditions. Therefore \eqref{sec3-eq7a} has the simple solution $\boldsymbol \mu(\theta) = \boldsymbol 0$ for all $\theta$. The stationary distribution is
\begin{equation}
p_{st}(\theta,\widetilde \bR) = \dfrac{1}{\sqrt{(2\pi)^n |\boldsymbol \sigma|}} \, \exp \left[ -\dfrac{1}{|\boldsymbol \sigma|} \sum_{i,j=1}^{n-1} |\boldsymbol \sigma_{ij}| \widetilde R_i \, \widetilde R_j\right] \label{sec3-eq8}
\end{equation}
where $|\boldsymbol \sigma|$ is the determinant of the covariance matrix and $|\boldsymbol \sigma_{ij}|$ is the cofactor. The stationary distribution is used to define a projection operator $P$, defined through its action over a generic function $f(\theta,\widetilde \bR,t)$
\begin{equation}
P f(\theta,\widetilde \bR,t) = p_{st}(\theta,\widetilde \bR) \int dR_1 \ldots \int dR_{n-1} f(\theta,\widetilde \bR,t) \label{sec3-eq9}
\end{equation}
Although not indicated explicitly, the integrals in \eqref{sec3-eq9} extend over the whole set of admissible values of $R_i$. The application of the projector to the PDF $p(\theta,\widetilde \bR,t)$ gives the marginal PDF for the phase variable
\begin{equation}
P p(\theta,\widetilde \bR, t) = p_{st}(\theta, \widetilde \bR) \, \hat p(\theta,t) \label{sec3-eq10}
\end{equation}
Together with the projector $P$ we introduce the complementary operator $Q$, such that $P+Q = \boldsymbol 1$, where $\boldsymbol 1$ is the identity operator. Applying the standard Mori--Zwanzig projection procedure (see \cite{Bonnin2013}) the reduced FPE for the marginal probability density function is obtained
\begin{equation}
 \dfrac{\partial P p}{\partial t}  = P L P p + P L e^{tQL} Q p + \int_0^t K(t-s) P p \; ds \label{sec3-eq11}
\end{equation}
where $L=L_0 + \varepsilon L_1 + \varepsilon^2 L_2$ and $K(t-s) = PL e^{(t-s)QL}QL$ is the memory kernel. The first term on the right hand side $PLP p(\theta,\widetilde R,t)$ is called the \emph{Markovian term}, since its contribution is completely determined by the initial value of the marginal PDF $\hat p(\theta,t)$. The second term is the \emph{noise term} and summarizes the contribution of the initial distribution of the amplitude variables $Q p(\theta,\widetilde \bR,0)$. If the initial value $p(\theta, \widetilde \bR, 0)$ is random, the function $Q p(\theta,\widetilde \bR,0)$ is a stochastic process. If the initial distribution is deterministic the noise term can be eliminated by a proper choice of the initial condition, because $Q p(\theta,\widetilde \bR,0)$ in the null space of $P$. The last term is the \emph{memory  term}, because it collects the dependence on the whole past history of the marginal PDF through the time integral. The non Markovian behavior of the reduced equation is a consequence of the attempt to describe the system evolution with a reduced number of variables \cite{Bonnin2013}.

The analysis of \eqref{sec3-eq11} is a very challenging problem, but it can be simplified with suitable approximations. In this work we adopt a short time approximation, assuming that the memory kernel decays so fast that we can drop the integral term. The analysis of the influence of the memory term is left to future works. Thus we consider the simplified FPE
\begin{equation}
\dfrac{\partial P p(\theta,\widetilde R,t)}{\partial t} \approx P L P p(\theta,\widetilde R,t) \label{sec3-eq12}
\end{equation}
Using equations \eqref{sec3-eq5} and \eqref{sec3-eq8} it is straightforward to find
\begin{subequations}
\begin{align}
P L_0 P p & =  - p_{st}(\theta,\widetilde \bR) \dfrac{\partial \hat p(\theta,t)}{\partial \theta} \\[1ex]
P L_1 P p & =  0\\[1ex]
P L_2 P p & =  - p_{st}(\theta,\widetilde \bR) \bigg\{ \dfrac{\partial }{\partial \theta}\bigg[ \dfrac{1}{2}  \sum_{i,j=1}^{n-1} \dfrac{\partial^2 a_{\theta}}{\partial R_i \partial R_j} E[\widetilde R_i \widetilde R_j] + \hat a_{\theta} \bigg ] \hat p(\theta,t) \bigg\} + \dfrac{1}{2} \dfrac{\partial^2}{\partial \theta^2} \bB_{\theta} \bB_{\theta}^T \hat p(\theta,t)
\end{align}\label{sec3-eq13}
\end{subequations}
Introducing \eqref{sec3-eq10} and \eqref{sec3-eq13} into \eqref{sec3-eq12} yields the reduced FPE for the marginal PDF $\hat p(\theta,t)$
\begin{align}
\nonumber \dfrac{\partial \hat p}{\partial t} & = - \dfrac{\partial }{\partial \theta} \left\{ \left[ 1 + \dfrac{\varepsilon^2}{2} \sum_{i,j=1}^{n-1} \dfrac{\partial^2 a_{\theta}}{\partial R_i \partial R_j} \, E[\widetilde R_i \, \widetilde R_j] + \varepsilon^2 \hat a_{\theta} \right] \hat p \right\} + \dfrac{\varepsilon^2}{2} \dfrac{\partial^2 }{\partial \theta^2} \left[ \bB_{\theta} \bB_{\theta}^T \hat p(\theta,t) \right] \label{sec3-eq14}
\end{align}
that corresponds to the reduced SDE for the phase variable
\begin{equation}
d \theta = \left[1 + \dfrac{\varepsilon^2}{2} \sum_{i,j=1}^{n-1} \dfrac{\partial^2 a_{\theta}}{\partial R_i \partial R_j} \, E[\widetilde R_i \, \widetilde R_j] + \varepsilon^2 \hat a_{\theta} \right] dt + \varepsilon \bB_{\theta} d\bW_t \label{sec3-eq15}
\end{equation}

With respect to the reduced phase model \eqref{sec3-eq2}, equation \eqref{sec3-eq15} includes an additional term of order $\mathcal{O}(\varepsilon^2)$. The additional frequency shift depends on the covariance of the amplitude deviation, that can be found solving \eqref{sec3-eq7b}. Taking the stochastic expectation on both sides of \eqref{sec3-eq15} and using the zero expectation property of It\^o SDE, the expected angular frequency of the reduced phase model is readily obtained
\begin{equation}
E\left[ \dfrac{d\theta}{dt} \right] = 1 + \dfrac{\varepsilon^2}{2} \sum_{i,j=1}^{n-1} E \left[\dfrac{\partial^2 a_{\theta}}{\partial R_i \partial R_j}\right] \, E[\widetilde R_i \, \widetilde R_j] + \varepsilon^2 E \left[\hat a_{\theta} \right]\label{sec3-eq16}
\end{equation}

\section{Example}
As an example of application, we consider a Duffing--van der Pol system with multiplicative noise, described by the state equations
\begin{subequations}
\begin{align}
d x = & \left[ y - \alpha\left(\dfrac{x^3}{3} - x \right) \right] dt + \varepsilon \,y \, dW_x \label{sec4-eq1a}\\[2ex]
dy = & \left( -x - \beta x^3 \right) dt + \varepsilon \, x\,  dW_y \label{sec4-eq1b}
\end{align}\label{sec4-eq1}
\end{subequations}
In the noiseless limit ($\varepsilon=0$), the analytic solution for this system cannot be expressed in terms of simple elementary functions \cite{kimiaeifar2009}. Thus we use numerical simulations to determine the limit cycle. The results is then used to compute the vectors $\bu_1(t)$, $\bu_2(t)$, $\bv_1(t)$, and $\bv_2(t)$, for both the orthogonal and the Floquet basis. Floquet vectors are computed using the formulas given in \cite{Bonnin2012}.
 \begin{figure}
\centering
 \includegraphics[width=60mm,angle=0]{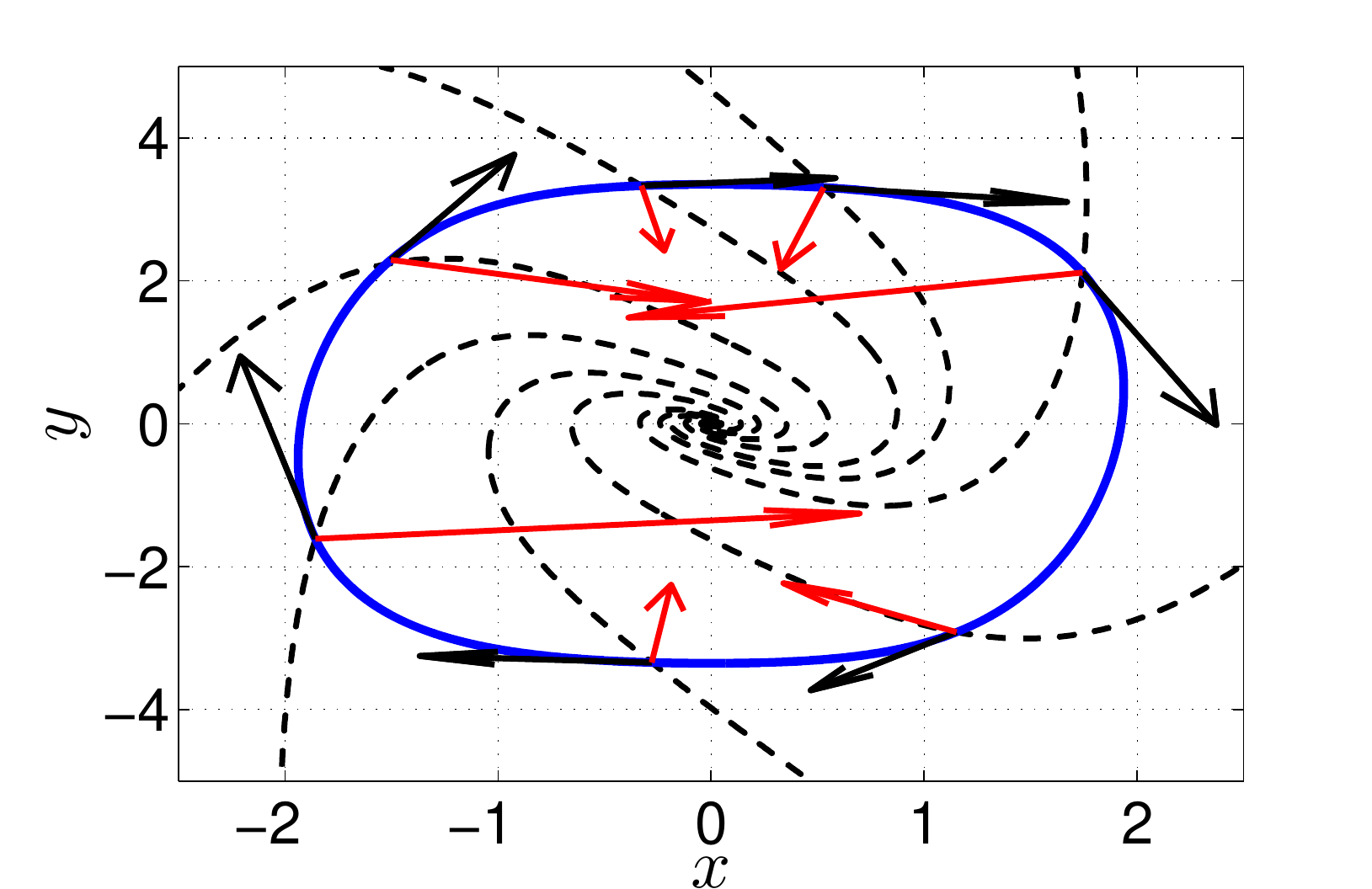}%
 \includegraphics[width=60mm,angle=0]{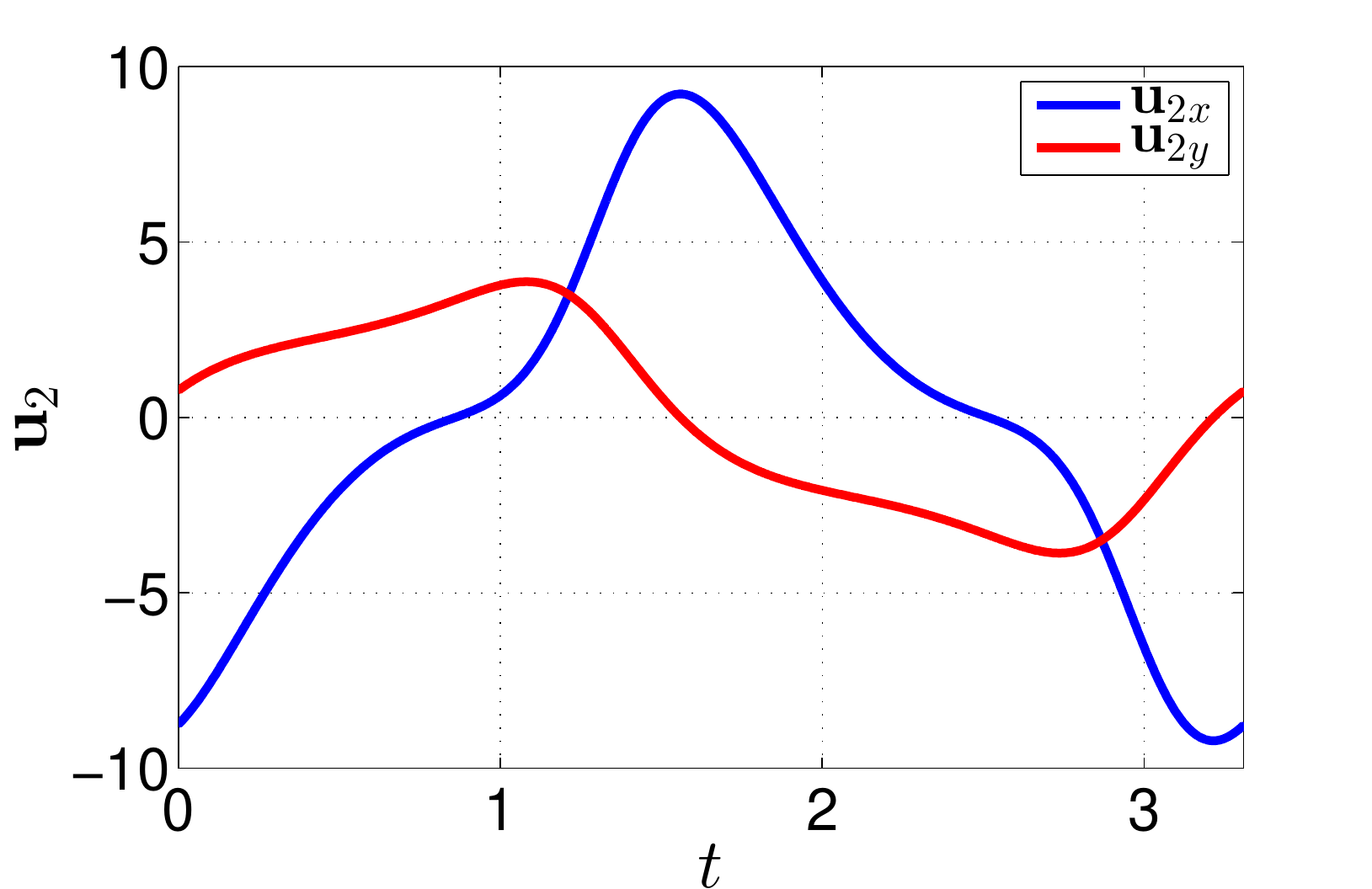}%
 \caption{Left: the limit cycle $\bx_s(t)$ of the Duffing--van der Pol oscillator \eqref{sec4-eq1} (blue line), the tangent vector $\bu_1(t)$ (black arrows) and the orthogonal vector $\bu_2(t)$ (red arrows) at different time instants. Some isochrons of the Duffing--van der Pol system are also shown for reference (black dashed lines). Right: time evolution for the components of $\bu_2(t)$ over one period. Parameters are: $\alpha=\beta=1$, $\varepsilon=0$.\label{figure1}}
 \end{figure}
 \begin{figure}
\centering
 \includegraphics[width=60mm,angle=0]{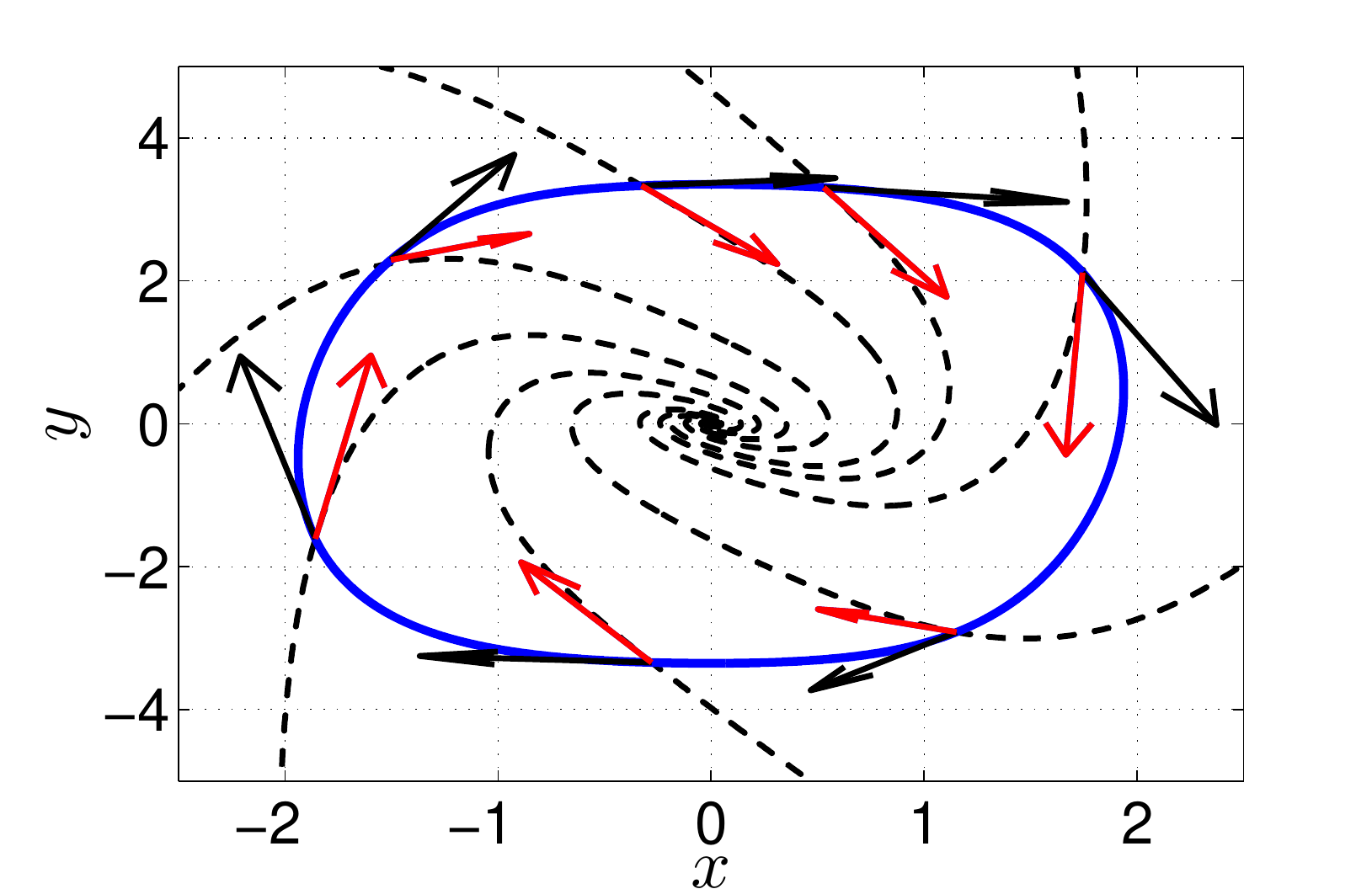}%
 \includegraphics[width=60mm,angle=0]{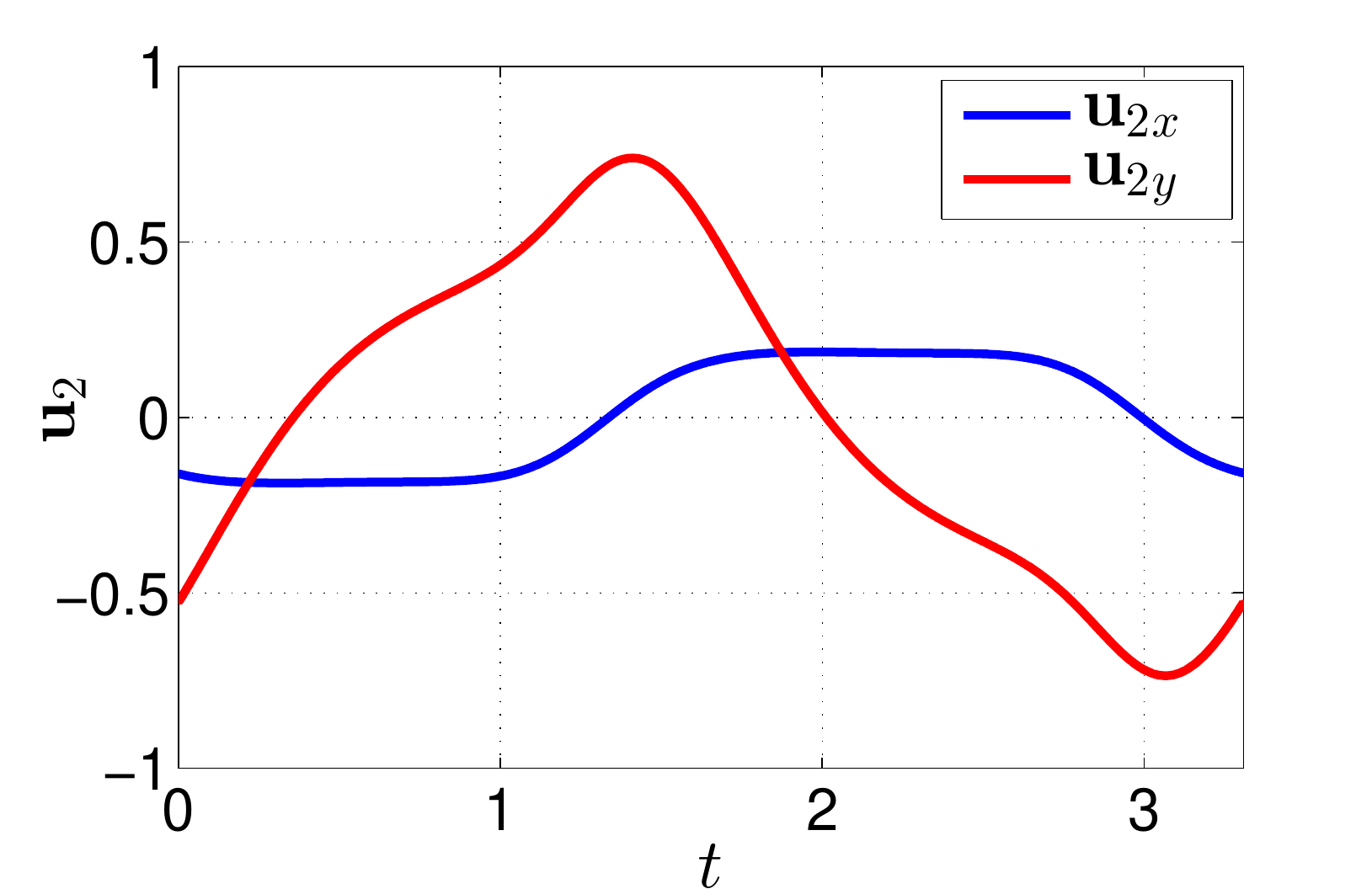}%
 \caption{Same as figure \ref{figure1} for the Floquet basis. \label{figure2}}
 \end{figure}
Figures \ref{figure1} and \ref{figure2} show the limit cycle, and the vectors $\bu_1(t)$,  $\bu_2(t)$ obtained with the orthogonal and the Floquet basis, respectively. Some isochrons (computed using the algorithm described in \cite{Izhikevich2007}, pag. 490) are also shown for reference. As expected, when Floquet basis is used the vector $\bu_2(t)$ is locally tangent to the isochron on the limit cycle. It is worth noting that for this example isochrons cannot be expressed in terms of elementary functions, therefore, an analysis based on isochrons would require their numerical computation (together with their first and second spatial derivatives, \cite{Bonnin2013}) for all points of the state space, a very demanding task.

\begin{figure}
\centering
 \includegraphics[width=60mm,angle=-0]{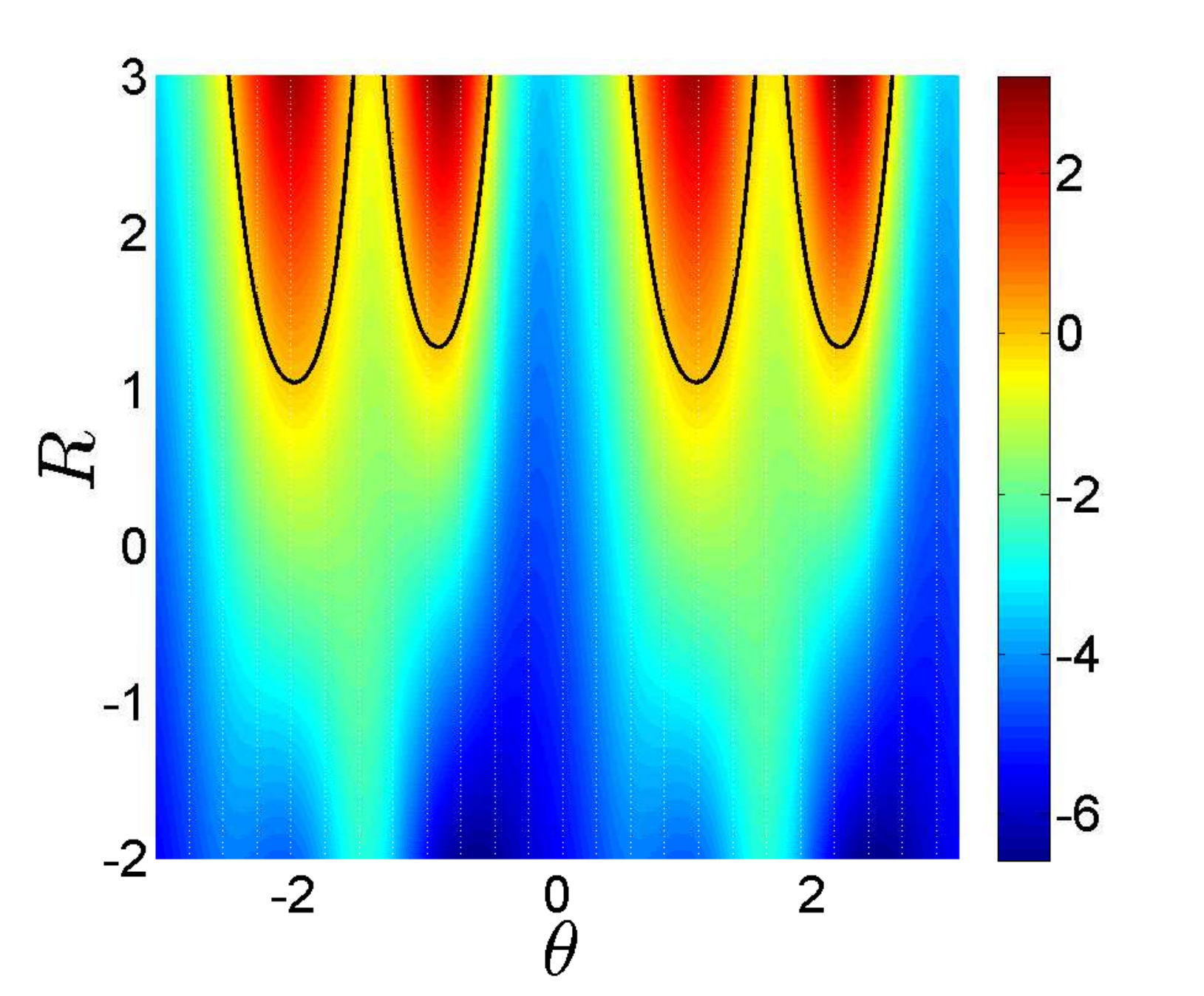}%
 \includegraphics[width=60mm,angle=-0]{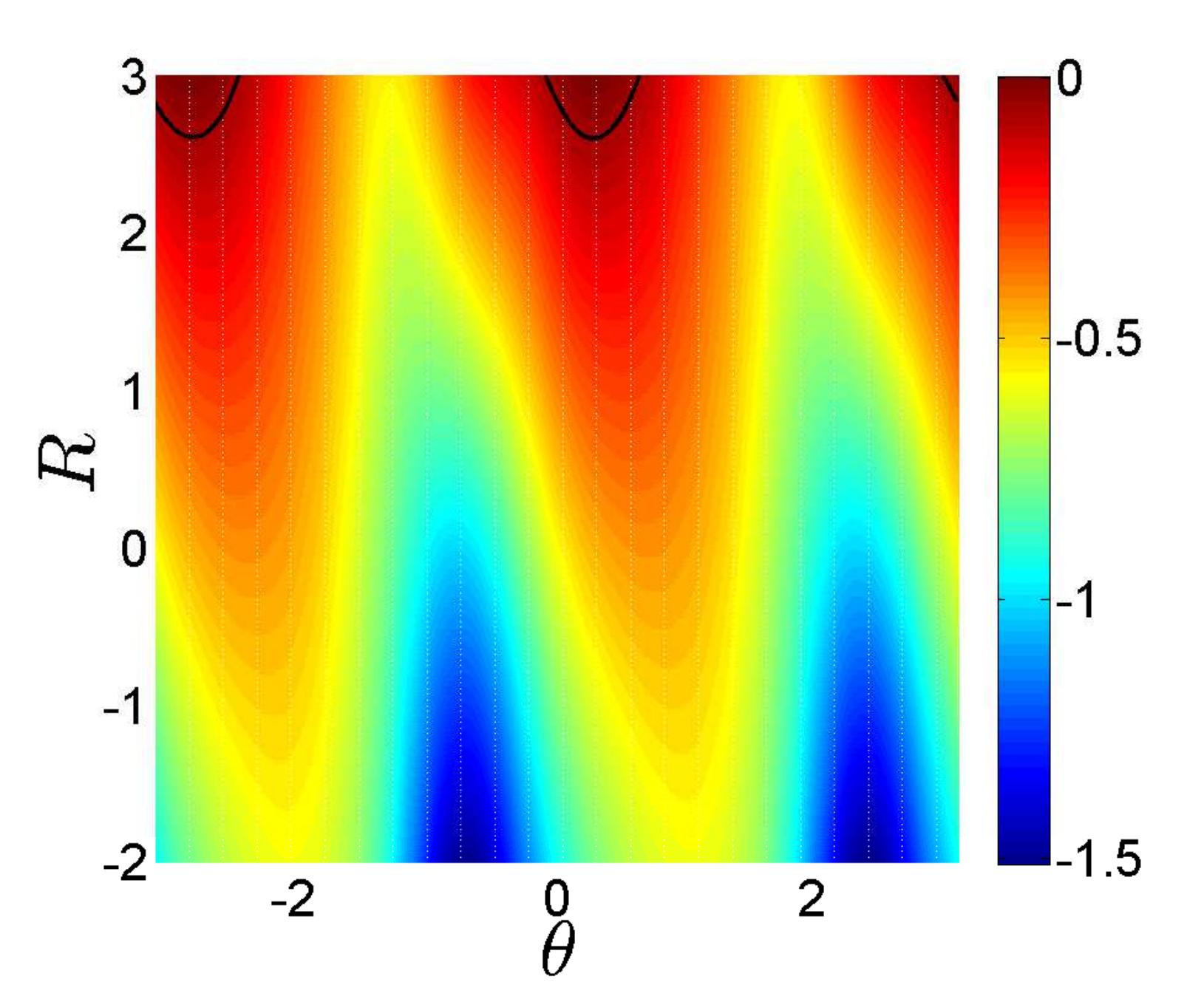}%
 \caption{Jacobian of the coordinate transformation \eqref{sec2-eq3}, for the Duffing--van der Pol oscillator. Colors indicate the value of the determinant, the thick black lines represent where the determinant is null, and thus the transformation is singular. Left: Jacobian with the orthogonal basis. Right: Floquet basis. Parameters are $\alpha=\beta=1$. \label{figure3}}
 \end{figure}

The limit cycle $\bx_s$ and the vectors $\bu_1$, $\bu_2$ are used to determine the coordinate transformation \eqref{sec2-eq3}. Figure \ref{figure3} shows the value of the Jacobian for the coordinate transformation of the Duffing--van der Pol oscillator. The thick black lines represent where the determinant is null, and thus the coordinate transformation to phase--amplitude variables is singular. It is evident that large values of the amplitude deviation (and thus of the noise intensity) can be reached before the amplitude--phase model is no longer valid.

\begin{figure}[h]
\centering
\includegraphics[width=70mm]{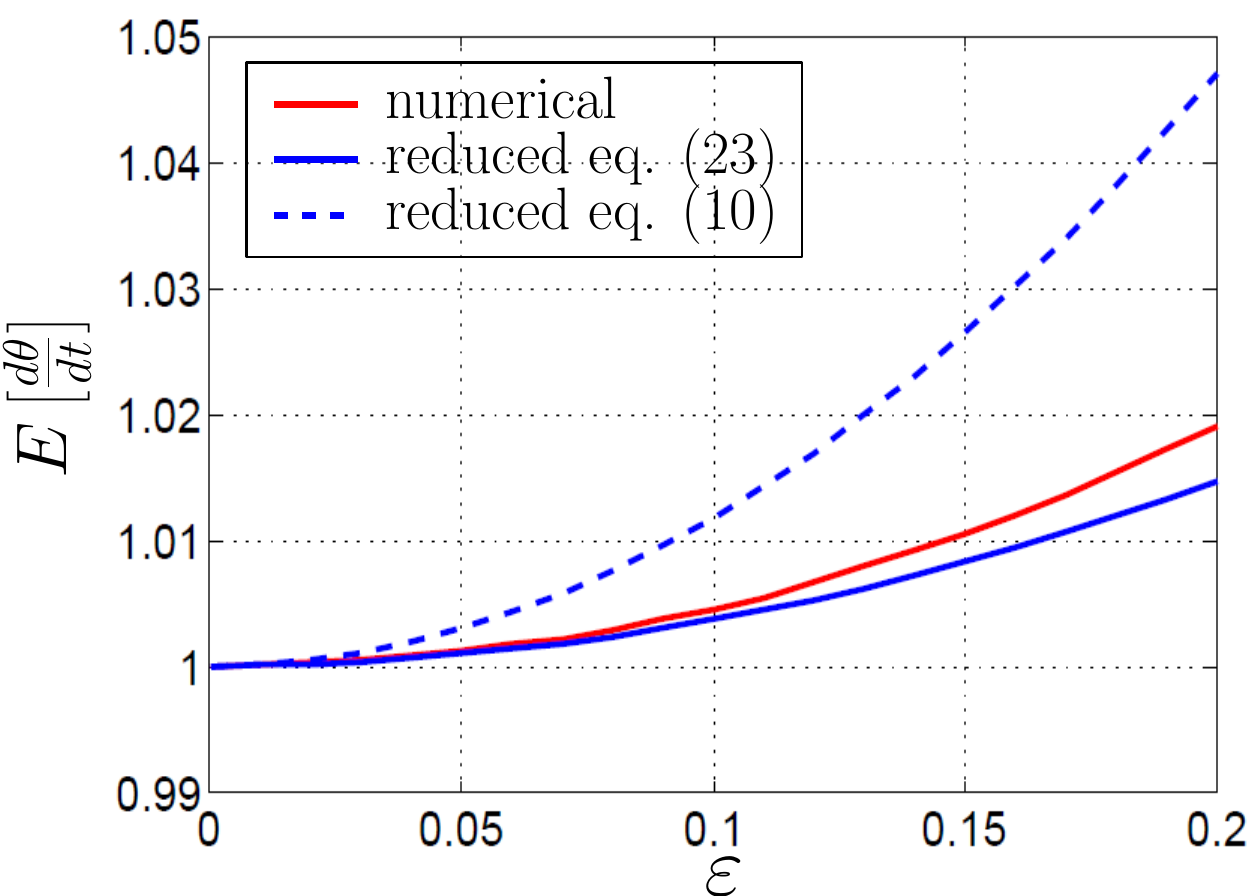}%
\caption{Expected normalized angular frequency vs the noisy intensity for the Duffing van der Pol oscillator \eqref{sec4-eq1}. Red line: numerical solution of the full phase and amplitude equations \eqref{sec2-eq4}. Blue dashed line: phase reduced model \eqref{sec3-eq2}. Blue solid line: improved phase reduced model \eqref{sec3-eq16}. Floquet basis is used, parameters are: $\alpha=\beta=1$.\label{figure4}}
\end{figure}

The vectors $\bu_1$, $\bu_2$, together with the covectors $\bv_1$, $\bv_2$ are used to compute the phase and amplitude equations \eqref{sec2-eq4}. For computational purposes, in particular for the derivatives that appear in \eqref{sec2-eq5}, it is convenient to consider a truncated Fourier series for $\bu_2$
\[ \bu_2(t) = \bu_{20} + \sum_{k=1}^N \bu_{2k}^a \cos (k \,\omega_0 t) + \bu_{2k}^b \sin (k \, \omega_0 t) \]
where $\omega_0$ is the fundamental frequency and $N$ is a sufficient high number of harmonics. The phase and amplitude equations \eqref{sec2-eq4} corresponding to the Duffing--van der Pol system \eqref{sec4-eq1} have been integrated numerically to determine the expected angular frequency, using both Euler--Maruyama and Milstein numerical integration schemes. The expected normalized angular frequency has been determined from the solution of the phase equation using
\[ E\left[\dfrac{d \theta}{dt}\right] = \dfrac{\theta(t_2)-\theta(t_1)}{t_2-t_1}\]
for $t_2\gg t_1$. This is justified under the hypothesis that the system is ergodic, so that ensemble averages coincide with time averages. The expressions for $\bx_s$, $\bu_1$, $\bu_2$, $\bv_1$ and $\bv_2$ also allow to derive and solve \eqref{sec3-eq7b}. Note that since the Duffing--van der Pol is a second order system, the variance $\sigma(\theta)$ is just a scalar.  The variance is then introduced into \eqref{sec3-eq16} to find the expected angular frequency given by the reduced phase oscillator model.
Figure \ref{figure4} shows the normalized expected angular frequency versus the noise intensity for the example under investigation. For simplicity, only Floquet basis is considered. The result obtained using numerical simulations (solid red line) is compared with the phase oscillator model given by \eqref{sec3-eq2} (dashed line) and with the improved model \eqref{sec3-eq16} (solid lines). It is clear that the proposed model \eqref{sec3-eq16} provides a much more accurate estimate of the expected angular frequency.

\section{Conclusions}

In this paper, a novel phase reduced model for nonlinear oscillators subject to white Gaussian noise has been presented. Starting from a recently developed description of the noisy oscillator dynamics in terms of amplitude and phase variables, a single scalar stochastic differential equation for the phase variable is given. The resulting phase oscillator model is analogous to the celebrated Kuramoto model.

The reduction procedure leading to the phase oscillator model is based on the use of projection operator techniques and stochastic calculus. It is shown that, with the help of Floquet theory, the phase variable obtained coincides in the vicinity of a limit cycle with the asymptotic phase defined using the concept of isochrons.

The phase equation shows a quadratic dependence of the phase dynamics on the noise intensity. It is shown that the expected angular frequency depends on the covariance matrix of amplitude variable. In particular, even in the case where the expected oscillation amplitude remains unchanged, the unavoidable amplitude fluctuations do modify the expected oscillation frequency.

As an example, a  Duffing--van der Pol oscillator is analyzed both theoretically and numerically. The validity of the proposed phase oscillator model is confirmed by the numerical analysis. It is also shown that the proposed model provides more accurate prediction that other models previously proposed in literature.

\end{document}